\documentclass[useAMS,usenatbib,onecolumn]{mn2e}

\title[A new technique for timing the double pulsar system]{A new
technique for timing the double pulsar system}
\author[Freire et al.]{P. C. C. Freire$^{1,2}$\thanks{E-mail: pfreire@naic.edu (PCCF)}, N. Wex$^{3}$, M. Kramer$^{3,4}$, D. R. Lorimer$^2$,
\newauthor M. A. McLaughlin$^2$,  I. H. Stairs$^{5,6}$, R. Rosen$^7$, and
A. G. Lyne$^4$
\\
$^{1}$Arecibo Observatory, HC3 Box 53995, Arecibo, PR 00612, USA\\
$^{2}$Department of Physics, West Virginia University, PO Box 6315
Morgantown, WV 26506-6315, USA\\
$^{3}$Max-Planck Institut f\"ur Radioastronomie, Auf dem H\"ugel 69, 53121 Bonn, Germany\\
$^{4}$University of Manchester, Jodrell Bank Centre for Astrophysics,
Alan-Turing Building, Oxford Road, Manchester, M13 9PL, UK \\
$^{5}$University of British Columbia, 6224 Agricultural Road
Vancouver, BC V6T 1Z1 Canada\\
$^{6}$ATNF, P.O. Box 76, Epping NSW 1710, Australia\\
$^{7}$NRAO, Green Bank, WV 24944-0002, USA
}
\begin{document}

\date{Draft, in preparation}

\pagerange{\pageref{firstpage}--\pageref{lastpage}} \pubyear{2008}

\maketitle

\label{firstpage}

\begin{abstract}
In 2004, McLaughlin et al.~discovered a phenomenon in the radio
emission of PSR~J0737$-$3039B (B) that resembles drifting
sub-pulses. The repeat rate of the sub-pulses is equal to the spin
frequency of PSR~J0737$-$3039A (A); this led to the suggestion that
they are caused by incidence upon B's magnetosphere of
electromagnetic radiation from A.  Here we describe a
geometrical model which predicts the delay of B's sub-pulses relative
to A's radio pulses. We show that measuring these delays is equivalent
to tracking A's rotation from the point of view of an hypothetical
observer located near B. This has three main astrophysical applications:
(a) to determine the sense of rotation of A
relative to its orbital plane; (b) to estimate where in B's
magnetosphere the radio sub-pulses are modulated and (c) to provide an
independent estimate of the mass ratio of A and B. The latter might
improve existing tests of gravitational theories using this system.
\end{abstract}


\section{Introduction}
\label{sec:intro}

The discovery of the double pulsar system PSR~J0737--3039
\cite{bdp+03,lbk+04} has led to important advances in the study of
radio pulsars and their associated phenomenology. The independent
determination of the orbit of both neutron stars, and the recent
measurement of five post-Keplerian parameters from the timing of
{\tt A} \cite{ksm+06} and a sixth one from eclipse observations of
{\tt A} \cite{bkk+08} has made this the most over-constrained system
known, allowing five tests of general relativity.

PSR~J0737$-$3039B (henceforth {\tt B}) has many unique features.  It
emits pulsed radio waves during most of the orbit, but it becomes much
brighter at two distinct orbital phases \cite{lbk+04}. Furthermore,
its pulse profile changes strongly with orbital phase and with time
\cite{bpm+05}. Though instructive in other ways, this behaviour has
made it difficult to measure times of arrival accurately and obtain a
precise estimate of the semi-major axis
of the orbit of {\tt B}, limiting the precision of our knowledge of
the pulsar mass ratio $R = m_A/m_B$ and the precision of some of the
tests of general relativity in this system (see, e.g., Kramer \&
Stairs 2008 for a recent review\nocite{ks08}). 

This paper is motivated by another unique feature of {\tt B}: for
certain orbital phases its radio emission is clearly modulated by
electromagnetic emission from {\tt A} \cite{mkl+04}; the phenomenon
superficially resembles drifting sub-pulses (e.g.~Lorimer \& Kramer
2005)\nocite{lk05}. In this work, we discuss
the {\em timing} of this phenomenon. In \S~\ref{sec:delays},
we present a theoretical calculation of the delay in the arrival at
the Earth of the sub-pulses of {\tt B} relative to the radio
pulses of {\tt A} responsible for the drift. We name this simply the
``response delay''. In \S~\ref{sec:comparison}, we discuss how we
can compare the predicted and measured response delays. In
\S~\ref{sec:discussion}, we highlight the astrophysical knowledge that
can be gained from the timing of the response delays.

\section{Calculating response delays}
\label{sec:delays}

\begin{figure*}
\setlength{\unitlength}{1in}
\begin{picture}(0,5.0)
\put(-3.0,0.0){\includegraphics{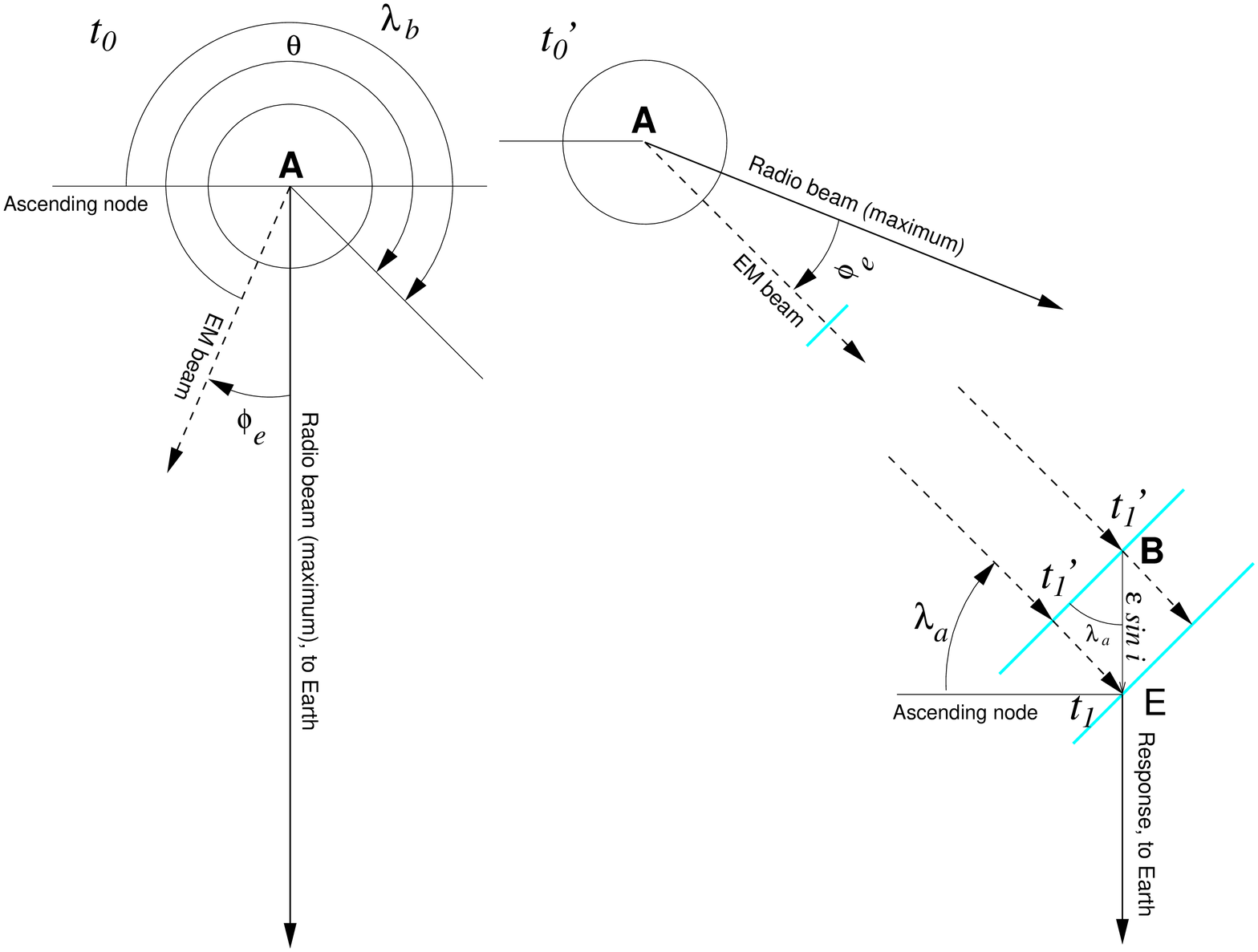}}
\end{picture}
\caption [] {\label{fig:rotation}
Sequence of events in {\tt A}, looking down the pulsar's spin axis.
We assume that the plane of the drawing is the orbital plane.
{\em Left}: at $t_0$ the radio beam is pointing at
the Earth (outside of the orbital plane), producing
a maximum of radio emission. At this time, {\tt A}'s EM beam is not
necessarily pointing at {\tt B}. {\em Center, top}: at $t_0'$ {\tt A} is
emitting the EM signal that will produce the response. Locally, the
wavefront (light line) is a plane perpendicular to the direction of
propagation of the EM signal. Between $t_0$ and $t_0'$, {\tt A} rotated
by $\theta = \lambda_b + \pi/2 - \phi_e$ if the pulsar is rotating
in the same sense of the orbit (clockwise here), or
$\theta = 2 \pi - (\lambda_b + \pi/2 - \phi_e)$ if the pulsar is rotating
in the opposite sense of the orbit. {\em Right, bottom} at time $t_1'$, the
EM wavefront arrives at {\tt B}'s position. Soon after that, at time
$t_1$, the wavefront arrives at point E, where the Earth-bound radio
emission of {\tt B} originates. This is located at a fixed
distance $\epsilon$ from {\tt B} on that pulsar's line of sight to
Earth, it is at a small distance from the orbital plane
$\epsilon \cos i$ (see text).
The arrival of the wavefront at point E
at time $t_1$ produces the response in {\tt B}'s emission, which then
follows to Earth. $\lambda_a$ is the longitude of {\tt A} as seen from
E at $t_1$; so $\lambda_b = \lambda_a + \pi$ is the longitude at which
{\tt A}'s EM beam transmits towards {\tt B}.}
\end{figure*}

\subsection{Time between emission of {\tt A}'s radio pulse and
{\tt B}'s response}
\label{sec:rotation}

The time between the emission of {\tt A}'s radio pulse and
{\tt B}'s response can be divided in three parts, which we discuss 
below.

\begin{enumerate}
\item The sub-pulses of {\tt B} are spaced by 22.7~ms,
which is the same as the spin period of {\tt A}
\cite{mkl+04}. Therefore, we can think of the maximum intensity of the
sub-pulses of {\tt B} (the ``response'') as being caused by
the impact upon {\tt B}'s radio emitting region of a co-rotating beam of
electromagnetic radiation from {\tt A}.
We henceforth refer to this as the EM beam. Since the EM beam
is not necessarily aligned in rotational phase with {\tt A}'s radio beam, 
as shown in Fig.~\ref{fig:rotation}, we introduce a phase offset 
$\phi_e$ between the two beams. We define {\tt A}'s zero longitude to
be the point where the radio emission towards the Earth reaches
maximum intensity. As we show later, one of the applications of our
model would be the direct determination of $\phi_e$.

One of the simplifying assumptions in Fig.~\ref{fig:rotation} is that
the spin axis of {\tt A} is perpendicular to the orbital
plane.  This has been suggested \cite{mkp+05} as the most
plausible explanation for the lack of change in the pulse profile of
{\tt A}; Ferdman (2008)\nocite{fer08} set an upper limit
of about $14^\circ$ for the misalignment between the orbital angular
momentum and the spin of A. We henceforth assume that the spin
axis of {\tt A} is indeed perpendicular to the orbital plane.
It is not know whether {\tt A} rotates in the same sense as the orbit
or in the opposite sense, but kinematic considerations \cite{bai88}
make the former more likely. We discuss this issue further in
\S~\ref{sec:comparison} and \S~\ref{sec:discussion}.

In Fig.~\ref{fig:rotation}, at time $t_0$ we see {\tt A}'s radio beam
pointing at the Earth, causing the radio emission to reach a
maximum. At time $t_0'$ the EM beam is pointing at a direction
$\lambda_b$, where it emits the signal that will later cause {\tt B}'s
response\footnote{Lower-case
subscripts refer to $\lambda$ computed for the emission-reception-Earth
triangle, while uppercase subscripts refer to lambda computed for a
line through the centre of mass of the binary system, O.
Here we have neglected aberration effects and the small angle between
AB and AE, as the corresponding time delays are below the accuracy
relevant for this paper.}. Between the two
events {\tt A} has rotated by an angle $\theta = \pm (\lambda_a -
\pi/2 - \phi_e)$ where $\lambda_a = \lambda_b - \pi$ is the longitude
of {\tt A} as seen from the location and time where the EM signal
modulated the radio pulse of {\tt B} (point E at time $t_1$, see
Fig.~\ref{fig:rotation}). The positive sign corresponds to a
clockwise rotation of {\tt A} as shown in Fig.~\ref{fig:rotation},
i.e., in the same sense of the orbit. The time between the two events
is given by
\begin{equation}
t_0' - t_0 = P_A \frac{\theta}{2 \pi},
\
\end{equation}
where $P_A$ is the spin period of {\tt A}.

\item After $t_0'$ the EM pulse of {\tt A} travels towards {\tt B}; at
time $t_1'$ it gets to {\tt B}'s exact position. The distance
between these two events is $r_{AB} (\lambda_a)$, this is calculated
in \S~\ref{sec:distance}. In the assumption above (that signal travels
at the speed of light) we have
\begin{equation}
t_1' - t_0'  = \frac{r_{AB}(\lambda_a)}{c}.
\end{equation}

\item What happens near {\tt B} is less certain. The simplest assumption is
that the EM signal is able to travel through {\tt B}'s magnetosphere
at the speed of light (this is possible e.g. if the EM signal consists of
high-energy photons) and modulates {\tt B}'s Earth-bound radio
emission at the point where it is being generated; we designate this
as ``E''.

We must keep in mind that this simplified assumption is not the only
possible case: a) the modulation could occur after B's radio signal is
produced, i.e., somewhere between point E and the Earth and b) more
generally, the chain of events from the emission of the EM signal at
{\tt A} ($t_{0}'$) to the modulation of the radio signal of {\tt B}
($t_1$) might not lie along a single null geodesic.
For the time being, we will quantify the simplest case only (see
Fig.~\ref{fig:rotation}).

Since E is where the Earth-bound radio emission is being generated, we
assume it is the point of the radio-emitting region of {\tt B}'s
magnetosphere that is in the line of sight from {\tt B} to Earth.
This only happens when {\tt B}'s radio pulse is ``on''. For
most of {\tt B}'s rotational cycle, when its pulse is ``off'', no part
of its radio-emitting region is pointing at Earth, and no response is
detected.

In our simplified model, we assume that the distance of E to {\tt B}
is a constant $\epsilon$ as a function of {\tt B}'s spin phase. Therefore, in
the reference frame of the orbital plane (with axes along the line of
nodes, perpendicular to the line of nodes and perpendicular to the
orbital plane), E is located at fixed coordinates (0, $\epsilon \sin
i$, $\epsilon \cos i$) relative to {\tt B}, where $i$ is the orbital
inclination of the system (see Fig.~\ref{fig:rotation}).

At $t_1'$, the EM wavefront from {\tt A} reaches {\tt B}. A small
amount of time later, at $t_1$, the same wavefront reaches E and
produces the response.
 From that Figure, we can see that
\begin{equation}
t_1 - t_1' = \frac{\epsilon}{c} \sin \lambda_a \sin i.
\end{equation}
Between $t_1'$ and $t_1$ the orbital motion of {\tt B} will make E move
by a small amount relative to the center of the binary, this will
affect the arrival time at E. We note,
however, that for slow pulsars $\epsilon/c$ is
expected to be of the order of 1 ms \cite{bcw91}. Assuming this is
true, a detailed calculation of the $\mathit O (\epsilon c^{-2})$
Doppler correction due to {\tt B}'s (and E's) motion shows that this
term is smaller than $1\,\mu \rm s$. Since the r.m.s. timing precision
of {\tt A} is about $10\,\mu \rm s$, and the timing precision for
{\tt B}'s response is likely to be worse, we will ignore all terms
with magnitude $<\,10\,\mu \rm s$.
\end{enumerate}
For the total time difference we obtain:
\begin{equation}
\label{eq:time}
t_1 - t_0 \equiv (t_1 - t_1') + (t_1' - t_0') + (t_0' - t_0)
= \frac{\epsilon}{c} \sin \lambda_a \sin i +
\frac{r_{AB}(\lambda_a)}{c} + P_A \frac{\theta}{2 \pi} + \mathit O ( \epsilon c^{-2}),
\end{equation}
which should be valid in any inertial reference frame using the
orbital parameters as measured in that frame.

\begin{figure*}
\setlength{\unitlength}{1in}
\begin{picture}(0,6)
\put(-2.4,0.0){\includegraphics{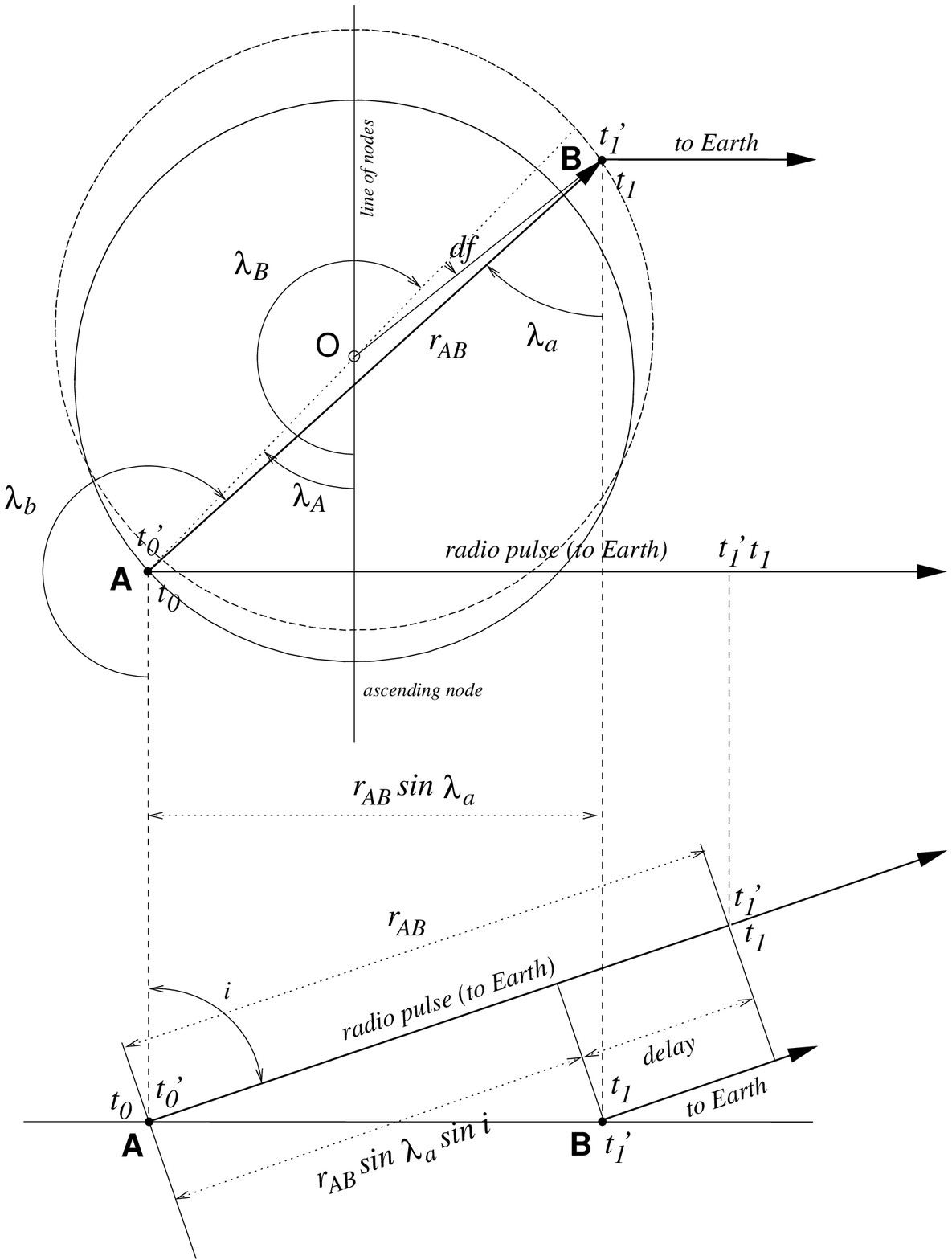}}
\end{picture}
\caption [] {\label{fig:travel_time}
Emission (at $t_0'$, on {\tt A}) and reception at the position of
{\tt B} (at $t_1'$) of the EM signal that causes {\tt B}'s responses. 
{\it Upper diagram:} events depicted in the orbital plane. O represents the
centre of mass of the system, orbital motion is represented as clockwise.
{\it Lower diagram:} events depicted on a plane that is perpendicular to
the orbital plane and the plane of the sky. This contains the line of sight
to the Earth and allows a representation of the response delays. The orbital
inclination $i$ of the PSR~J0737$-$3039 binary system is much closer to
90$^\circ$ than depicted in this figure \cite{ksm+06};
we represent it as being substantially lower for clarity. The
difference in the locations of $t_0,t_0'$ and $t_1',t_1$ is too small
to be discerned at this scale. In this figure, the line of periastron
just happens to coincide with the line of nodes, with $\omega_A = 180^\circ$,
$\omega_B = 0^\circ$.}
\end{figure*}

\subsection{Response delay}
\label{sec:delay}

We now calculate the delay between the reception at the SSB of {\tt
  A}'s radio pulses and {\tt B}'s responses,
$\Delta (\lambda_a)$. To do this, we will start by adding to $t_1 - t_0$ the
difference in the ranges\footnote{We use the term ``range'' to indicate
the distance of a given event from the center of mass of the binary
projected along the direction to the SSB {\em as seen at the SSB}.}
at which these events occur, $z_1 - z_0$ divided by $c$; this is the
so-called ``R{\o}mer delay''.

As in the case of $t_1 - t_0$ (eq.~\ref{eq:time}), $z_1 - z_0$ can
also be described as the sum of three terms, $(z_1 - z_1') + (z_1' -
z_0') + (z_0' - z_0)$:
\begin{enumerate}
\item  The difference in range between the events at $t_0$ and $t_0'$
  is given by:
\begin{equation}
z_0' - z_0 = (t_0' - t_0) v_{l,A} = P_A \frac{\theta}{2 \pi} v_{l,A},
\end{equation}
where $v_{l,A}$ is the velocity of {\tt A} relative to the centre of mass,
projected along the line of sight to Earth (\S \ref{sec:distance}).

\item In the lower diagram of Fig.~\ref{fig:travel_time}, we
see that the difference in range between the events happening
at $t_1'$ and $t_0'$ is given by:
\begin{equation}
z_1' - z_0' = - r_{AB}(\lambda_a) \sin \lambda_a \sin i.
\end{equation}
\item Ignoring the motion of E between $t_1'$ and $t_1$, we have
\begin{equation}
z_1 - z_1' = - \epsilon.
\end{equation}
\end{enumerate}
The delay is therefore given by:
\begin{equation}
\Delta_r (\lambda_a) = t_1 - t_0  + \frac{z_1 - z_0}{c}
= \frac{r_{AB}(\lambda_a) - \epsilon}{c} \left( 1 - \sin \lambda_a
\sin i \right) + P_A' \frac{\theta}{2 \pi}
+ \mathit O ( \epsilon c^{-2}),
\label{eq:basic}
\end{equation}
where the subscript ``r'' indicates that we are taking into account
the R{\o}mer delay only. In this equation we use $P_A' = P_A (1 +
v_{l,A}/c)$ to represent the spin period of {\tt A} with the classical
Doppler correction due to its orbital velocity. This equation is valid
at the SSB if we use the orbital parameters as measured there.

In Appendix A we present a detailed calculation of the relativistic
contribution to the total response delay $\Delta$, known as the
``Shapiro delay'' ($\Delta_s$). For the range of orbital phases where
we observe {\tt B}'s responses $\Delta_s$ changes $4\, \mu \rm
s$, i.e., smaller than the r.m.s. timing precision of {\tt A}. We can
therefore assume $\Delta = \Delta_r + \Delta_s \simeq \Delta_r$ for
the remainder of this paper.

\begin{figure*}
\setlength{\unitlength}{1in}
\begin{picture}(0,4)
\put(-2.5,0.0){\includegraphics{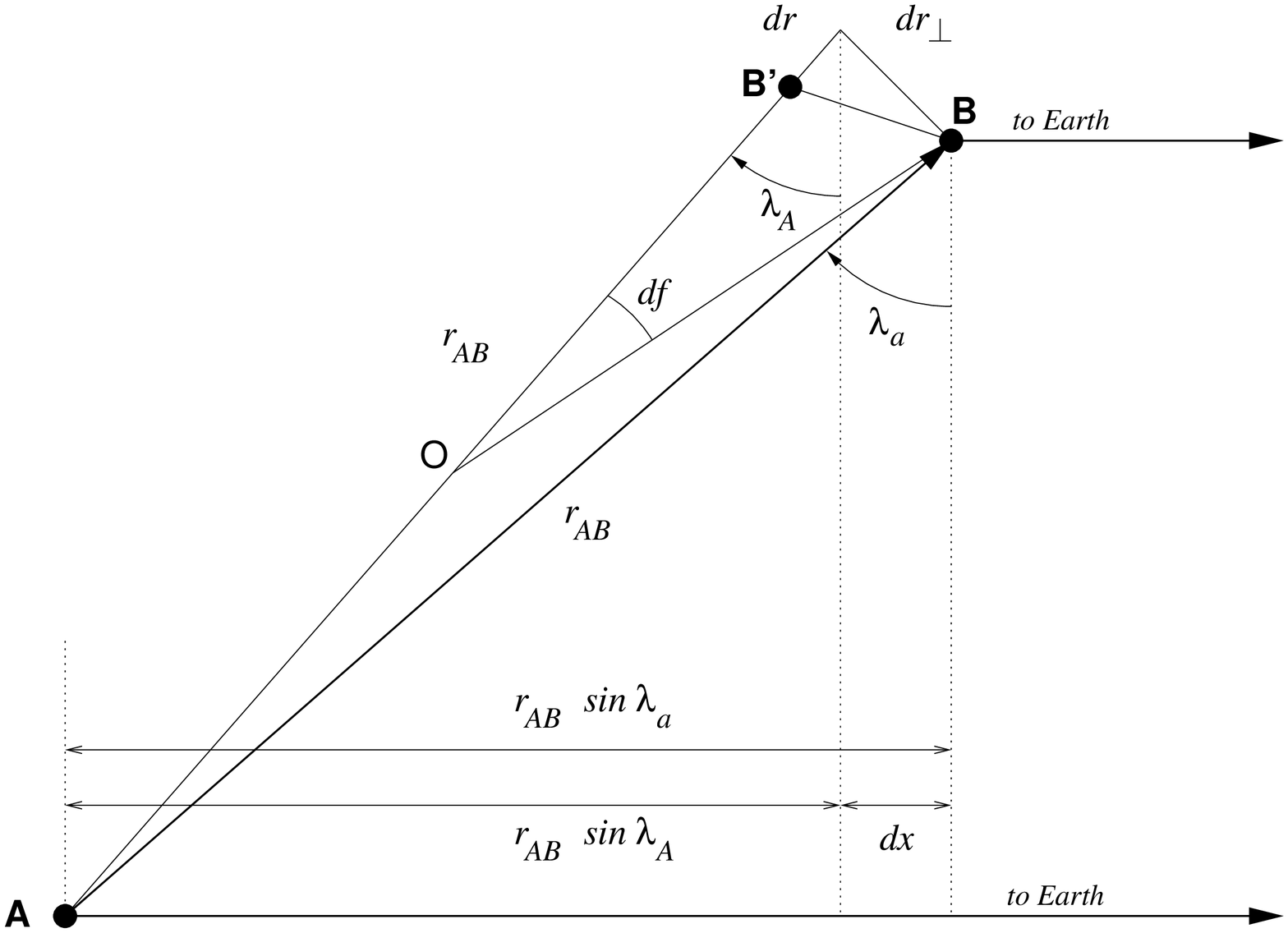}}
\end{picture}
\caption [] {\label{fig:blowup}
Detail on the motion of {\tt B} in the orbital plane of the system.
O represents the centre of mass of the system.}
\end{figure*}

\subsection{Distance between {\tt A} at transmission and {\tt B} at response}
\label{sec:distance}

Having calculated the response delay as a function of $\lambda_a$ and
the separation between the two pulsars, $r_{AB} (\lambda_a)$, we now
calculate these two quantities as a function of $\lambda_A$ at
$t_0'$. The reason for this is that $\lambda_A$ can be calculated
precisely for any given time using the equations in Damour \& Deruelle
(1985,1986)\nocite{dd85,dd86}, furthermore, this quantity determines
the instantaneous configuration of the system.

In the equations that follow, we will
only use the Newtonian terms to $\mathit O (c^{-2})$. A detailed
calculation shows that the $\mathit O (c^{-3})$ Newtonian terms are much
smaller than the r.m.s. timing precision of {\tt A}. This means that
they can be ignored for or present purposes.

At $t_0'$, the separation of pulsar {\tt I} ({\tt A} or {\tt B}) from
the centre of mass is given (see, e.g., Roy 1988\nocite{roy88}) by
\begin{equation}
d_I = a_I \frac{1 - e^2}{ 1 + e \cos f},
\label{eq:delay_0}
\end{equation}
where $a_I$ is the semi-major axis of its orbit, $e$ is the
orbital eccentricity and the angle $f$ is the true anomaly of the system at
$t_0'$ ($f = \lambda_I - \omega_I$, where
$\omega_I$ is the longitude of periastron of pulsar {\tt I} at $t_0'$).
The components of a pulsar's velocity along the radial (i.e., away from
the centre of mass) and transverse directions are given \cite{roy88} by
\begin{eqnarray}
\dot{d_I} & = & v_I e \sin f \label{eq:rdot},\\
d_I \dot{f} & = & v_I ( 1 + e \cos f) \label{eq:fdot},
\end{eqnarray}
where
\begin{equation}
v_I = \frac{2 \pi}{P_b} \frac{a_I}{\sqrt{1 - e^2}},
\end{equation}
is the transverse velocity of pulsar {\tt I} at quadrature ($f = \pi /2$) and
$P_b$ is the orbital period of the binary. The velocity of that pulsar relative
to Earth is given by (e.g., Green 1985)\nocite{gre85}:
\begin{equation}
v_{l,I} = v_I \sin i \left( \cos \lambda_I  + e \cos \omega_I \right).
\end{equation}
Calculating this velocity for {\tt A} we can immediately quantify the
Doppler correction term for $P_A$ in eq. \ref{eq:basic}.

To calculate the instantaneous distance between the pulsars ($d_{AB}$), we
replace $a_I$ by $a = a_A + a_B$ in eq. \ref{eq:delay_0}. During the
time it takes {\tt A}'s signal to cross this distance ($d_{AB}/c$),
{\tt B} is moving. Because the EM signal is moving in the radial
direction, only the radial component of {\tt B}'s motion ($dr$) will
affect the travel time between the two pulsars. During the cross time
$d r \simeq \dot{d_B} d_{AB} / c$, i.e., the distance between the
events at $t_0'$ and $t_1'$ is
\begin{equation}
r_{AB}(\lambda_A) = d_{AB} + dr = d_{AB} \left(1 + \frac{v_B}{c} e \sin f \right) + \mathit{O} \left( c^{-2} \right);
\label{eq:dr}
\end{equation}
the perpendicular motion of {\tt B} between $t_0'$ and $t_1'$ is
\begin{equation}
d r_{\perp} \equiv d_B df = d_B \dot{f} (t_1' - t_0') = v_B (1 + e \cos f) \frac{r_{AB}(\lambda_A)}{c}.
\label{eq:r_perp}
\end{equation}
Looking at Fig.~\ref{fig:blowup} we can see that
\begin{equation}
\lambda_a = \lambda_A + \frac{d r_{\perp}}{r_{AB}(\lambda_A)} = \lambda_A + \frac{v_B}{c} (1 + e \cos f)\label{eq:lambda}.
\end{equation}
In the derivation above we made the approximation that $\lambda_a$ is
the longitude of {\tt A} as seen from {\tt B} at $t_1'$.
The longitude of {\tt A} as seen from point E at $t_1$ is larger by a
very small amount, $\epsilon \sin i \cos \lambda_a / r_{AB}$.

Noting that $dx = d r_{\perp} \cos \lambda_A$,
\begin{equation}
r_{AB}(\lambda_A) \sin \lambda_a = r_{AB}(\lambda_A) \sin \lambda_A + d r_{\perp} \cos \lambda_A.\label{eq:rsinl}
\end{equation}
With this result we can re-write eq. \ref{eq:basic} as a function of
$\lambda_A$:
\begin{equation}
\Delta (\lambda_A) = \frac{r_{AB}(\lambda_A) - \epsilon}{c} \left( 1
  - \sin \lambda_A \sin i \right) - \frac{dr_{\perp}}{c} \cos
\lambda_A \sin i  + P_A' \frac{\theta}{2 \pi} + \mathit O ( \epsilon c^{-2}).
\end{equation}
Using eqs. \ref{eq:delay_0}, \ref{eq:dr} and \ref{eq:r_perp}, we can
re-write this as a function of a Keplerian term, $K(\lambda_A)$, which
can be calculated from known orbital parameters and the system's
geometry, and the unknown quantities
$\epsilon$ and $\theta$:
\begin{eqnarray}
\Delta (\lambda_A)& = & K(\lambda_A) - \frac{\epsilon}{c} \left( 1 -
\sin \lambda_A \sin i \right ) + P_A' \frac{\theta}{2 \pi}  + \mathit O ( \epsilon c^{-2})\label{eq:fundamental} \\
K(\lambda_A) & = & \frac{a}{c} (1 - e^2) \left[ \frac{1 - \sin
    \lambda_A \sin i}{1 + e \cos f} \left( 1 + \frac{v_B}{c} e \sin f
  \right)  - \frac{v_B}{c} \cos \lambda_A \sin i \right] +  \mathit{O} \left( c^{-3} \right).
\label{eq:K}
\end{eqnarray}
The $K(\lambda_A)$ term is the time difference, measured at the SSB,
between the events that occurred at $t_1'$ and $t_0'$ in the reference
frame of the binary; it is by far the largest contribution to the
response delay.

\section{Comparing measurements of response delays with predictions}
\label{sec:comparison}

So far, we have just made a theoretical calculation of the response
delays, $\Delta (\lambda_A)$. We now discuss what we can learn from
actually measuring such delays.

\subsection{Measuring absolute delays}
\label{sec:single_orbit}

If we measure $\Delta (\lambda_A)$ for several $\lambda_A$ in a
single orbit and subtract $K(\lambda_A)$, we can measure the delays
that are not {\em a priori} predictable:
\begin{equation}
\Delta (\lambda_A) - K(\lambda_A) = \frac{\epsilon}{c}
\left( \sin \lambda_a \sin i - 1 \right)
\pm P_A' \left( \frac{\lambda_a}{2 \pi}- \frac{\phi_e}{2 \pi} -
\frac{1}{4} \right).
\label{eq:remains}
\end{equation}
In order to measure $\Delta (\lambda_A)$, we have
to specify a particular radio pulse of {\tt A} (in principle the
closest to the time of emission of the EM signal that caused that
particular response) and subtract its (barycentric) time from the
response's barycentric time. However, because we have no prior
knowledge of the terms in eq. \ref{eq:remains}, there is some
ambiguity in the choice of the pulse that is closest to that time of
emission.

An experimental method that does away with the issue of the
choice of a pulse of {\tt A} and even the measurement of $\Delta
(\lambda_A)$ entirely would proceed along the following lines:

\begin{enumerate}
\item {\tt B}'s responses are particularly noticeable in an intensity
  gray-scale plot of time versus spin phase of {\tt B} as
  presented by \cite{mkl+04}. In such a plot,
  we can determine the precise spin phase corresponding
  to each response; the precise method to accomplish this and the
  attainable precision will be discussed elsewhere.
\item In the next step, we convert this phase into a barycentric
time. This can be achieved using the software package {\sc
  tempo}\footnote{http://www.atnf.csiro.au/research/pulsar/tempo/}.
  The result corresponds to an event that occurred at $t_1$ in the
  reference frame of the binary.
\item Subtracting $K(\lambda_A)$, we obtain a second barycentric time
  that nearly corresponds to {\tt A}'s emission of the EM pulse ($t_0'
  + \epsilon / c \sin \lambda_A \sin i) $ in the reference frame
  of the binary). At this time we calculate $\lambda_A$ from the
  binary's ephemeris.
\item Subtract the nearest time at which the rotational phase of {\tt
  A} is zero. This time should be the equivalent of $\Delta
  (\lambda_A) - K(\lambda_A)$. By fitting this to eq. \ref{eq:remains}
  we should be able to determine $\epsilon$, $\phi_e$ and
  the direction of the spin of {\tt A}.
\end{enumerate}

With data from different orbits, we can also get a sense
for the stability of $\phi_e$ and $\epsilon$. Thus far we have assumed
that $\epsilon$ (the height above {\tt B} where the response
is produced) is a constant as a function of {\tt B}'s rotational
phase. It is possible, however, that $\epsilon$ varies with {\tt B}'s
spin phase. This would also introduce a secular variation of the
$\epsilon$ observed for any spin phase of {\tt B} due to the pulsar's
geodetic precession. If that is the case, we could map the
height of E above {\tt B} as a function of the precession
phase. Confusion with $K(\lambda_A)$ can be avoided because, despite
the fact that both terms vary with $1 - \sin \lambda_A \sin i$, only
the latter varies with $(1 + e \cos f)^{-1}$.

\subsection{Delay variations at a constant $\lambda_A$}
\label{sec:constant_longitude}

The response delays $\Delta (\lambda_A)$ will vary in time for any
given $\lambda_A$ because of the apsidal motion $\dot{\omega}$ changes
$f$, which causes a change in $K(\lambda_A)$:
\begin{equation}\label{eq:delta_1}
\delta_{t_2 - t_1} (\lambda_A)  \equiv  \Delta_{t =
  t_2}(\lambda_A) - \Delta_{t = t_1}(\lambda_A) = \frac{a}{c} (1 -
e^2) \left( 1 - \sin \lambda_A \sin i \right) [ j(t_2) - j(t_1)],
\end{equation}
where $t_1$ and $t_2$ are two epochs at which the longitude $\lambda_A$
occurs and
\begin{equation}
j(t) = \left( 1 + \frac{v_B}{c} e \sin f \right)
\left( 1 +e \cos f \right)^{-1}.
\end{equation}
Other terms of order larger than $c^{-3}$ cancel out in
the subtraction because of the constant $\lambda_a$, and the same
happens for the terms dependent on $\theta$ and $\epsilon$\footnote{The
  latter term is likely to cancel even if the main assumption in \S
  \ref{sec:rotation} is not correct, i.e., it only requires that any
  terms dependent on $\epsilon$ vary with $\lambda_a$ in a repeatable
  way.}. The variation of the Doppler correction to $P_A$
are $<\,1\,\mu$s and can therefore be ignored.

The $\delta_{t_2 - t_1} (\lambda_A)$ are important because, being
due solely to a variation of $K(\lambda_A)$, they are proportional
to $a$. The maximum variation occurs
between the time when $\lambda_A$ coincides with periastron
($f = 0$, occurring at $t = t_p$) and apastron ($f = \pi$, occurring at
$t = t_a$, which for PSR~J0737$-$3039 is about 10.65 years later):
\begin{equation}\label{eq:delta_2}
\delta_{t_a - t_p} (\lambda_A) = \frac{2 a e}{c} \left( 1 - \sin \lambda_A \sin i \right).
\end{equation}
In a single 2003 observation, McLaughlin et al.~(2004)
observe {\tt B}'s responses from
$\sim 195^\circ < \lambda_B < \sim 225^\circ$, i.e., occurring shortly
before the magnetosphere of {\tt B} eclipses {\tt A} at $\lambda_B = 270^\circ$. This corresponds to $\sim 15^\circ < \lambda_A < \sim
45^\circ$.  At these extremes we have $\delta_{t_a - t_p}
(\lambda_A) = 0.3814(3) \rm s$ and $0.1508(1) \rm s$.
The uncertainty in this prediction is entirely due to the uncertainty
of $a$ (which results entirely from the uncertainty in the
measurement of $a_B$); it represents 0.5 -- 1.5\% of a
rotation of {\tt A}; this means that we should be able to keep track
of the response delays with very high confidence.

The variations themselves are certainly measurable, since they are
equivalent to many rotations of {\tt A}. This means that we can always
make an independent measurement of $a$. Depending on the precision and
number of measurements of $\Delta (\lambda_A)$, the value of $a$
derived in this fashion might be more precise than the present value
derived from timing. In that case, we can improve our knowledge of the
mass ratio:
\begin{equation}
R = \frac{a \sin i}{c x_A} - 1,
\label{eq:mass_ratio}
\end{equation}
where $x_A$ is the projected semi-major axis of {\tt A}'s orbit, in
light seconds; this quantity is directly and very precisely measured
from the timing of {\tt A}. This equation (and also
eqs. \ref{eq:delta_1} and \ref{eq:delta_2}) show that, to determine $R$
from $a$ and $x_A$ we need to know $\sin i$. We could in principle
determine $\sin i$ independently from comparing measurements of
$\delta_{t_2 - t_1} (\lambda_A)$ made at different orbital
longitudes. However, most theories of gravitation
predict the Shapiro ``s'' term to be the same as $\sin i$
\cite{dt92,wil93}; therefore such an independent determination is not
likely to be a useful test of gravitation. For this reason, we can use
the extremely precise estimate of $\sin i$ from $s$ in the
estimate of $R$.

Finally, we remark that there is a small difference between the
intrinsic and observed semi-major axes of the pulsar's orbits
that is caused by aberration effects. Kramer et al. (2006) estimate
that for {\tt A} these are of the order of $10^{-6} x_A$, i.e., about
$1\,\mu \rm s$; this is similar to the
level of precision for the measurement of $x_A$. It is highly unlikely
that we will measure $a$ to a similar precision, so this should not be
a problem for the determination of $R$ in eq.~\ref{eq:mass_ratio}.
For {\tt B}, Kramer et al. estimate that the effect of aberration
should be of the order of $10^{-4} x_B$. It might be possible that we
measure $x_B = a \sin i / c - x_A$ to better than the necessary level
of precision. However, to be able to measure the aberration directly,
we must be able to measure $x_B$ from timing with this level
of accuracy, so that the difference between the two values becomes
evident. The intrinsic uncertainty of the $x_B$ obtained from timing is
about $10^{-3} x_B$ \cite{ksm+06}, one order of magnitude larger
than the effect of aberration. Therefore, unless the direct timing
of {\tt B} improves by more than one order of magnitude, the effects
of aberration won't be separately measurable.

\section{Implications and prospects}
\label{sec:discussion}

In this paper we have calculated the delay between the
radio pulses of {\tt A} and the modulated radio pulses of {\tt B},
$\Delta (\lambda_A)$, assuming a simple scenario for that
modulation, i.e., that it happens at the point, E, where the radio
emission of {\tt B} is being produced and that the
modulating signal is traveling from {\tt A} to E at the speed of
light. Things could be more complicated, particularly considering that
this trajectory must intersect the magnetosphere of {\tt B}.

We will present details of the measurements of $\Delta (\lambda_A)$
and their timing analysis elsewhere. If the response delays obey the
equations presented above in a consistent manner, that will validate
our model. In that case, the measurement of these
delays relative to the radio pulses of {\tt A} can provide us with new
astrophysical information.

First, a determination of the sense of {\tt A}'s rotation
relative to its orbit, something never achieved for any other pulsar,
would introduce fundamental constraints on binary
evolution scenarios for this pulsar, as well as improved constraints
on {\tt B}'s supernova kick. Finding that the orbital angular momentum
is anti-aligned with {\tt A}'s angular momentum would require a very
large supernova kick \cite{bai88}. Previous studies
(e.g., Willems et al. 2006, Stairs et al. 2006)\nocite{wkf+06,std+06}
predict instead that the kick that produced
{\tt B} was rather small, therefore the angular momenta should be
aligned. A confirmation of this alignment, combined with the small
angle between the momenta, would introduce stringent constraints on the
magnitude of the kick. Furthermore, we would know the sign of
the expected relativistic spin-orbit contribution to $\dot{\omega}$
\cite{ds88}. It is probable that this contribution will be measured in
the near future \cite{kw09}; that would
allow an estimate of {\tt A}'s moment of inertia. The moment of
inertia of {\tt A}, together with the well-determined mass of the
pulsar, will introduce fundamental constraints on the equation of
state for dense matter \cite{ls05}.

Second, by measuring $\epsilon$, or by introducing upper limits to it,
we might be able to locate the region where the EM signal from
{\tt A} is modulating the radio signal from {\tt B} and relate it to
the region where we expect the radio emission of {\tt B} is being
generated. Thus far, the best location of pulsar radio
emission comes from the interpretation of multi-frequency polarimetric
pulse profiles in light of a relativistic version
of the the familiar rotating vector model \cite{bcw91}; the results
indicate an emission height of a few hundred km. If the signal is
being modulated as its being generated, then $\epsilon$ should be of
the order of a few ms and therefore a measurable quantity. If
$\epsilon$ is significantly larger, then the modulation is happening
after {\tt B}'s radio signal is generated.
 
Third, we will make an independent measurement of the orbital
separation of the two pulsars, $a$. If the timing of {\tt B}'s 
responses is precise enough, this might give us a more precise
measurement of the mass ratio of {\tt A} and {\tt B}, which would
increase the precision of some of the previous tests of general
relativity carried out in this binary system. This is a key input
parameter for the general tests on conservative gravity theories
outlined in Kramer~\&~Wex~(2009)\nocite{kw09}.

In the ideal case that we can track the response times well, one
might think of their reception at the Earth as being equivalent to
having a radio telescope at an altitude $\epsilon$ above {\tt B}
tracking {\tt A}'s rotation and then relaying the results live to
Earth. This analogy highlights how fortunate we are to have a
phenomenon like {\tt B}'s responses.



\section*{Acknowledgments}

PCCF, DRL and MAM acknowledge support from a WVEPSCoR research challenge
grant held by the WVU Center for Astrophysics. MAM is an Alfred P. 
Sloan Fellow. Pulsar research at UBC is supported by an NSERC
Discovery Grant. IHS acknowledges sabbatical support from the ATNF
Distinguished Visitor program.

\label{lastpage}

\section*{Appendix A}
\label{sec:relativistic}

Including post-Newtonian corrections, the time taken for a photon to propagate 
from the point of emission ${\bf x}_0$ to ${\bf x}$ in the gravitational field
of a $N$-body system is given by
\begin{equation}\label{eq:tnbody}
  t - t_0 = \frac{|{\bf x} - {\bf x}_0|}{c} + \frac{2G}{c^3}\sum_{i=1}^{N}
            m_i\ln\left[ \frac{r_{0i} + r_i + R}{r_{0i} + r_i - R} \right] \;,
\end{equation}
where $m_i$ is the mass of the $i$-th body located at ${\bf x}_i$, $r_{0i}
\equiv |{\bf x}_i-{\bf x}_0|$, $r_i \equiv |{\bf x} - {\bf x}_i|$, and $R
\equiv |{\bf x} - {\bf x}_0|$ \cite{bru91}. The first term,
$|{\bf x} - {\bf x}_0| / c$, is the R{\o}mer term used in the main text. The
second term, the relativistic correction, is the sum of
the Shapiro delays caused by the individual bodies in the $N$-body system.

We now apply equation (\ref{eq:tnbody}) to the signal propagation discussed in
this paper (see figures \ref{fig:rotation} and
\ref{fig:travel_time}). Neglecting terms smaller than a few $\mu$s one
finds for the signal propagating from {\tt A} to Earth at distance $D$
\begin{equation}
  \Delta_{s}^{({\rm A\oplus})} \simeq
    \frac{2Gm_A}{c^3}\ln\left[\frac{D}{\delta_{\rm R}}\right] +
    \frac{2Gm_B}{c^3}\ln\left[\frac{2D}{d_{AB}(1+\cos\lambda_{a,\oplus})}\right] \;.
\end{equation}
$\delta_{\rm R}$ ($\ll d_{AB}$) is the emission height of the radio signal in
the magnetosphere of {\tt A} and $\lambda_{a,\oplus}$ is the angle between the
direction to Earth and the direction to pulsar {\tt A} as seen from
{\tt B}. For the signal propagating from {\tt A} to the point E to
sufficient accuracy%
\begin{equation}
  \Delta_{s}^{({\rm AE})} \simeq  
    \frac{2Gm_A}{c^3}\ln\left[\frac{d_{AB}}{\delta_{\rm EM}}\right] +
    \frac{2Gm_B}{c^3}\ln\left[\frac{2d_{AB}}{\epsilon(1+\cos\lambda_{a,\oplus})}\right] \;.
\end{equation}
$\delta_{\rm EM}$ ($\ll d_{AB}$) is the emission height of the EM signal that
triggers the sub-pulse emission at E. Finally, for the sub-pulse signal emitted
at E the Shapiro delay is given by
\begin{equation}
  \Delta_{s}^{({\rm E\oplus})} \simeq
    \frac{2Gm_A}{c^3}\ln\left[\frac{2D}{d_{AB}(1 - \cos\lambda_{a,\oplus})}\right] +   
    \frac{2Gm_B}{c^3}\ln\left[\frac{D}{\epsilon}\right] \;.
\end{equation}
The contribution of the Shapiro delays to $\Delta$ is therefore given by
\begin{eqnarray}\label{eq:deltashap}
  \Delta_{s} & = &\Delta_{s}^{({\rm AE})} + \Delta_{s}^{({\rm E\oplus})}
  - \Delta_{s}^{({\rm A\oplus})} \nonumber \\
& = &
  \frac{2Gm_A}{c^3}\ln\left[\frac{2}{1 + \sin i\sin\lambda_A}\right] +
  \frac{2Gm_A}{c^3}\ln\left[\frac{\delta_{\rm R}}{\delta_{\rm EM}}\right] +
  \frac{4Gm_B}{c^3}\ln\left[\frac{d_{AB}}{\epsilon}\right] \;,
\end{eqnarray}
where we have used $\cos\lambda_{a,\oplus} = -\sin
i\sin\lambda_A$. The second term of
equation (\ref{eq:deltashap}) is constant and, to sufficient accuracy, can be
absorbed in $\phi_e$. The third term of equation (\ref{eq:deltashap}) changes
along the eccentric orbit with an amplitude of just 2 $\mu$s, and therefore can
be absorbed into $\phi_e$ as well. In principle the first term of equation
(\ref{eq:deltashap}) can change quite significantly along the orbit, as $\sin i
\simeq 1$ in the double pulsar \cite{ksm+06}. However, as mentioned above, the
sub-pulses are only observed in the range $\sim 15^\circ$ $<$ $\lambda_A$ $<$
$\sim 45^\circ$. Across this interval the first term of equation
(\ref{eq:deltashap}) changes by less than 4 $\mu$s. Consequently,
Shapiro delays can be ignored in the calculations of this paper.

\end{document}